\documentclass[conference]{IEEEtran}
\usepackage{amsmath,amssymb,amsfonts}
\usepackage{algorithmic}
\usepackage{algorithm}
\usepackage{array}
\usepackage[caption=false,font=normalsize,labelfont=sf,textfont=sf]{subfig}
\usepackage{textcomp}
\usepackage{stfloats}
\usepackage{url}
\usepackage{verbatim}
\usepackage{graphicx}
\usepackage{cite}
\usepackage{tikz}
\usepackage{pgfplots}
\pgfplotsset{compat=1.18} 
\usepackage{xcolor}
\usepackage{microtype}
\usepackage{mathtools}
\usepackage{pgf-pie}
\usepackage{enumitem}

\usetikzlibrary{chains}

\def\BibTeX{{\rm B\kern-.05em{\sc i\kern-.025em b}\kern-.08em
    T\kern-.1667em\lower.7ex\hbox{E}\kern-.125emX}}
\usepackage{braket}
\usepackage{hyperref}
\usepackage{physics}

\begin{document}
\bibliographystyle{IEEEtran}

\title{A Web-based Software Development Kit for Quantum Network Simulation}

\author{Stephen DiAdamo and Francesco Vista \\ \{sdiadamo, fvista\}@cisco.com \\ Cisco Quantum Lab, Germany \vspace{-5mm}}


\maketitle

\begin{abstract}
  Quantum network simulation is an essential step towards developing applications for quantum networks and determining minimal requirements for the network hardware. As it is with classical networking, a simulation ecosystem allows for application development, standardization, and overall community building. Currently, there is limited traction towards building a quantum networking community---there are limited open-source platforms, challenging frameworks with steep learning curves, and strong requirements of software engineering skills. Our Quantum Network Development Kit (QNDK) project aims to solve these issues. It includes a graphical user interface to easily develop and run quantum network simulations with very little code. It integrates various quantum network simulation  engines and provides a single interface to them, allowing users to use the features from any of them. Further, it deploys simulation execution in a cloud environment, offloading strong computing requirements to a high-performance computing system. In this paper, we detail the core features of the QNDK and outline the development roadmap to enabling virtual quantum testbeds.
\end{abstract}

\begin{IEEEkeywords}
Quantum networks, quantum communication, simulation, emulation, virtual lab, quantum software.
\end{IEEEkeywords}

\section{Introduction}

Quantum computing and quantum networking will bring new applications that are not possible to implement using conventional technology alone. These include enhancements in security, speedups in computing, and super-classical coordination of devices \cite{singh2021quantum}. Quantum systems have additional properties that give them more power than purely classical counterparts. These properties are superposition, unclonability, and entanglement. Superposition allows intrinsic randomness and additional power in computing, unclonability implies an arbitrary quantum state cannot be copied, which is useful for security and privacy, and entanglement between systems allows for enhancement in coordination. To make use of these additional properties in a network setting, quantum systems need to be sent between the communicating parties. For that, there needs to be networks and networking protocols in place.

\begin{figure}
    \centering
    \begin{tikzpicture}[
scale=0.95, 
every node/.style={transform shape},
start chain,
node distance=5mm]

\node[align=center, scale=0.92, on chain] (s1) at (0, 0) {\textbf{Design}};
\node[align=center, scale=0.92, on chain] (s2) {\textbf{Simulation}};
\node[align=center, scale=0.92, on chain] (s3)  {\textbf{Lab}};
\node[align=center, scale=0.92, on chain] (s4) {\textbf{Testbed}};
\node[align=center, scale=0.92, on chain] (s5) {\textbf{Deploy}};

\path [blue, out=70,in=110, <-, line width=0.35mm] (s1) edge (s2);
\path [red, out=70,in=110, <-, line width=0.35mm] (s1) edge (s3);
\path [green!80!black, out=70,in=110, <-, line width=0.35mm] (s1) edge (s4);
\path [orange, out=70,in=110, <-, line width=0.35mm] (s1) edge (s5);
\draw [black, ->, line width=0.2mm] (s1) -- (s2);
\draw [black, ->, line width=0.2mm] (s2) -- (s3);
\draw [black, ->, line width=0.2mm] (s3) -- (s4);
\draw [black, ->, line width=0.2mm] (s4) -- (s5);

\end{tikzpicture}
    \caption{The life-cycle for network technologies.}
    \label{fig:network-lifecycle}
\end{figure}
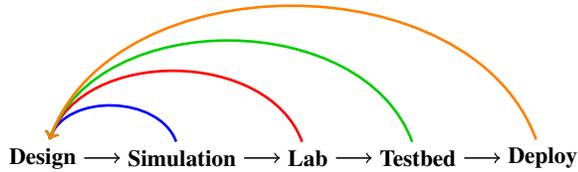

For any novel networking technology, there is a general life-cycle of development. As depicted in Fig.~\ref{fig:network-lifecycle}, for a new networking problem, the life-cycle begins by designing a solution, at first, under optimal conditions. This means there is no consideration of noise or external entities affecting the solution. Once the solution is designed, it is benchmarked and analyzed under a simulation setting. In simulation, additional challenges that the solution needs to overcome to be robust can be introduced in a controlled way. If the solution has flaws, it is updated in a design stage and then again evaluated in simulation. After this stage, the solution is taken to a lab environment with real devices. Here, noise is still controllable, but less so than in a simulated environment, and so new hurdles may appear that need to be overcome when testing in a lab. 

If additional hurdles appear, the solution is again updated to solve the new issues. The updated solution will again need to be simulated and analyzed before it can be moved into a testbed environment. In the testbed environment, noise cannot be controlled, as a testbed is usually deployed in the same environment as the target network. Again new issues may arise that affect the robustness of the protocol, and updates may need to be made. The process repeats from step one until the solution can finally be deployed in production. 

The important takeaway is that simulation appears early on and many times in the network technology life cycle. Therefore, there is a strong requirement for well developed, feature-rich, network simulation tools as they enable deploying novel network technologies. Now, when we consider quantum networks, we are in a very early stage, and many of the technologies that need to be developed have not yet left the design stage. Therefore, the need for quantum network simulation is paramount to successfully deploying production quantum networks. 

Developing quantum network technologies and protocols is non-trivial; they are generally not straightforward extensions of classical counterparts. To develop quantum network protocols, the quantum aspects that give quantum systems additional power also make it difficult to prepare and move the systems intact through the network. The simulation tools that exist in classical networking, therefore, will not be enough to validate quantum networking technologies and protocols and additional tools specific to quantum are needed.

Indeed, there has been a strong effort to develop quantum network simulation tools \cite{aji2021survey, bassoli2021quantum}. These simulators generally exist as Python-based frameworks in which a user defines a network topology, sets the associated network and hardware parameters, develops the protocol logic, and then executes a protocol over various parameters to benchmark and validate their protocol. An issue arises in that existing frameworks have many overlapping features and are structured in a similar way. Moreover, the simulation tools that exist are not universally adopted by the quantum network community since they generally 1) Do not provide any graphical interfaces to aid in building the simulation; 2) Require a strong level of knowledge regarding software development; 3) Need a strong knowledge of quantum networks to begin with; and; 4) Need a fairly powerful computer to execute bulk simulations. These together, from the authors' perspectives, are keeping the quantum networking community small and exclusive. 

In this work, we present our project, the QNDK, to unify the quantum network simulation community and solve these mentioned issues. It includes a graphical network visualization tool to easily build and run quantum network simulations. It exposes protocol code from an ever-expanding set of common quantum networking protocols to allow new users to start from meaningful examples. Further, it integrates various quantum network simulation backends providing a single interface, allowing users to switch between them and use the unique features of each without having to re-implement their simulations. The interface provides helpful tools to minimize the amount of code that needs to be written, even so far as a ``no-code" protocol development tool. Since the QNDK works as a web-based platform, it prepares a simulation which is sent to a cloud environment that runs the simulation and sends back the simulation results, offloading costly computations to a powerful cloud cluster. The user can use that data for benchmarking and protocol validation, achieving the same goal as current engines but with vastly less effort and complexity.

Overall, the goal of the QNDK project is to not only aid in simulation development, but eventually be used to deploy test-bed quantum networks in the field. The goal is to ``code once and deploy anywhere", meaning one should be able to run either simulation or execute the logic in a testbed without having to change the code. In this paper, we outline our current development status and detail our  roadmap toward enabling virtual quantum networks.

\section{The Quantum Network Development Kit}

In this section, we review the key features of the Quantum Network Development Kit (QNDK) and show an example demonstration.

\subsection{QNDK Main Features}

The QNDK aims to solve the core issues discussed in the introduction. It brings a graphical user interface and cloud execution to quantum network simulation to reduce the complexity of building and running complex simulations. The key features are the following:

\begin{enumerate}[wide]
    \item \textbf{Drag and drop interface for network topology}. This allows quick deployment of a network topology and the setting of the physical hardware parameters. The user can see an overview of the network components, easily importing hardware parameters from real systems and having a bird's-eye view of the overall network. This feature solves the problem of having to develop the topology with code.
    \item \textbf{Protocol development}. Part of the QNDK is that it provides the basis for developing new protocols from existing protocols. A user can access existing protocols and see their step-by-step instructions to make modifications or adapt from previous solutions to build new solutions. Protocol modification is done in the interface. 
    \item \textbf{Parallel and sequential assignment of protocol logic}. To make the network perform a protocol, the network nodes need to be configured to execute logic. There can be many protocols occurring simultaneously and some protocols require multiple sequential stages. Our interface is programmed to simply assign protocols to nodes to run in parallel or in series. This allows complex and rich simulation execution.
    \item \textbf{Changing the simulation engine}. Different simulation engines offer different features. The QNDK simplifies changing the engine by providing protocol logic in various engines, the currently supported being QuNetSim \cite{diadamo2021qunetsim} and NetSquid. \cite{coopmans2021netsquid}. A user simply changes a field in the interface and the application adapts accordingly. This allows users to have the best of all engines, developing a variety of simulations depending on their simulation needs.
    \item \textbf{Cloud execution of simulations}. Once the simulation is configured using the interface, the simulation can be sent up to a cloud server where it is compiled into a single simulation file. The simulation file executes on a server and so the user does not need to install any software locally. The file can also be downloaded to run and edit locally if desired.
    \item \textbf{Sharing and Exporting}. Once a simulation has been developed, a user can export the simulation configuration and code and share it. Another user can import the simulation and take over from where they left off, allowing for simple collaboration. We aim to make real-time collaborative editing a feature in the future to simplify this further.
\end{enumerate}

\subsection{Demonstration}

\begin{figure}[h]
    \centering
    \includegraphics[width=\columnwidth]{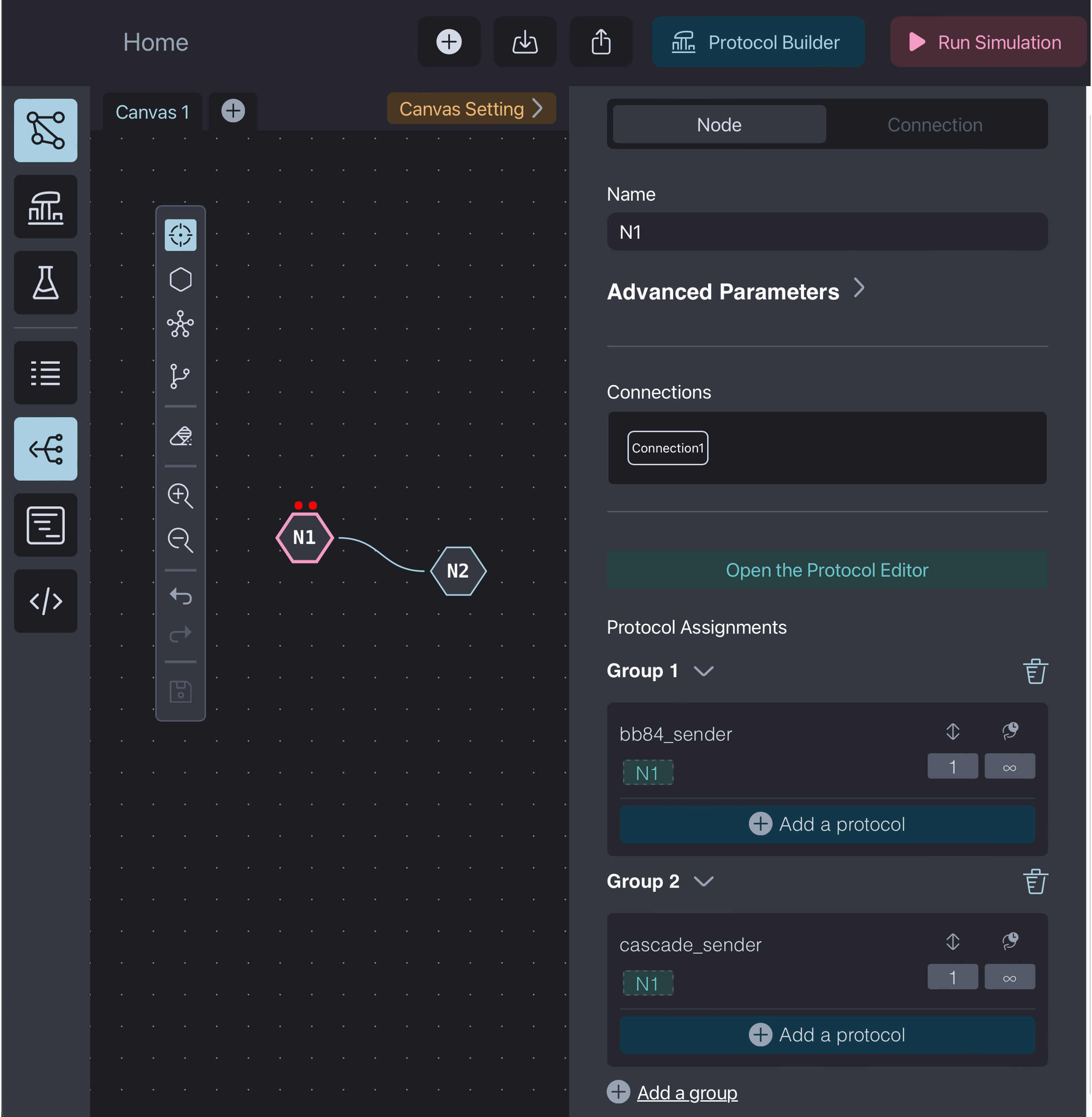}
    \caption{Topology creation and protocol assignment interface.}
    \label{fig:topology}
\end{figure}

\begin{figure}[h]
    \centering
    \includegraphics[width=\columnwidth]{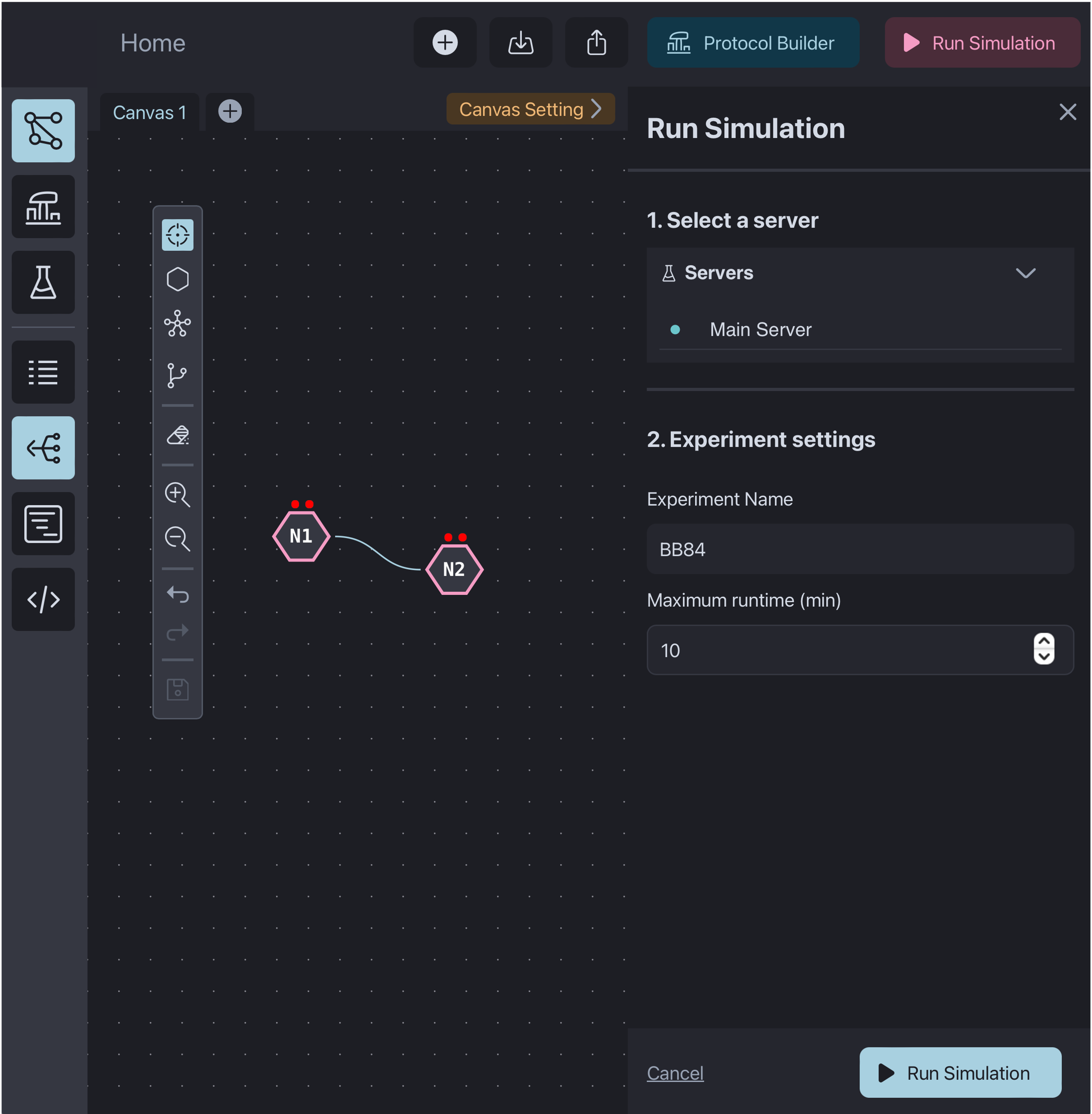}
    \caption{Run simulation interface.}
    \label{fig:run_sim}
\end{figure}

We demonstrate the QNDK's features by using a sample simulation based on BB84 QKD. We emphasize that the QNDK is built for general purpose quantum network simulation, not just for simulating QKD networks. We use QKD here only to demonstrate the features of the QNDK, using a well known multi-stage quantum network protocol. We detail the step-by-step flow to deploy the simulation.

\begin{enumerate}[wide]    
    \item \textbf{Topology creation}. The first phase is to create the network topology and the hardware details using the topology canvas and parameter inputs in the interface. Fig.~\ref{fig:topology} shows a two-node topology of quantum nodes. Each quantum node is characterized by various attributes such as quantum memory slots, hardware quality parameters (e.g., $T_1$ and $T_2$ times), and quantum source-related metrics like fidelity and qubit emission frequency. These parameters can be configured through the interface, accessed via the \textit{Advanced Parameters} menu. Additional settings related to connection such as fiber length, noise level, and signal loss can be set via the \textit{Connection} menu.
    \item \textbf{Protocol assignment}. Once the topology has been created, the protocols can be assigned. Protocols are assigned party by party. The first step of the QKD protocol is to distribute the key material and then perform the error correction step, so we construct two protocol groups. In Fig.~\ref{fig:topology}, node N1 is assigned the roles for the sender side of the protocol, so it is assigned \texttt{bb84\_sender} and \texttt{cascade\_sender}, while node N2 is designated as the receiver, and runs the \texttt{bb84\_receiver} and \texttt{cascade\_receiver} functionalities. A visual cue, denoted by a red dot atop each quantum node, denotes that the protocols have been assigned to the nodes, as depicted in Fig.~\ref{fig:run_sim}.
    \item \textbf{Run the simulation}. Following the protocol assignment, the simulation can be executed. Before running the simulation, one selects the server to send the simulation job to and then sends it for execution via the \textit{Run Simulation} button. The results of the simulation appear in the \textit{Experiments} tab, and they can be downloaded or viewed directly in the browser upon completion.
\end{enumerate} 

\section{Development Roadmap and Vision}

The QNDK is an ongoing project and the next steps for the project are the following. Firstly, the entire project will be made open source once it is in a state where we can accept community contributions. Next, we are working on simplifying developing protocols, and are working on a no-code interface for generating quantum networking protocols.

The vision for the QNDK is to act as a gateway to quantum network testbeds. This means users deploy their simulations not to the simulation server as we have shown, but rather to a network testbed. To achieve this, layers of compilers are needed to take high-level simulation code down to hardware instruction. The goal is for users to write their protocols once, simulate and validate them in a simulated network, and then send them to a physical quantum network to field test. 

Further, we aim to use the QNDK to deploy quantum networks. By integrating physical devices as emulators, we can create virtual quantum networks that can be deployed via a process similar to the one used in software-defined networks. This further goes in the direction of ``digital twins" for quantum networks, leveraging real-time information of the network to mimic the performance of the application.

Overall, there is a large scope of opportunity to develop software for quantum network simulation and deployment. Our QNDK project is a starting point, and we hope by open-sourcing the project, we can encourage the unification of the community, lowering the entry barrier and building a platform for future quantum networks.

\bibliographystyle{IEEEtran}

\end{document}